\documentclass[pre,twocolumn,aps,eqsecnum]{revtex4}
\usepackage{amssymb,amstext,amsmath}
\usepackage{graphicx}
\begin{document}

\title{Thermodynamic dislocation theory: Bauschinger effect} 
\author{K.C. Le, T.M. Tran} 
\affiliation{Lehrstuhl f\"ur Mechanik - Materialtheorie, Ruhr-Universit\"at Bochum, D-44780 Bochum, Germany}

\date{\today}

\begin{abstract}
The thermodynamic dislocation theory developed for non-uniform plastic deformations is used here to simulate the stress-strain curves for crystals subjected to anti-plane shear-controlled load reversal. We show that the presence of the positive back stress during the load reversal reduces the magnitude of shear stress required to pull excess dislocations back to the center of the specimen. There, the excess dislocations of opposite signs meet and annihilate each other leading to the Bauschinger effect.     
\end{abstract}

\maketitle

\section{Introduction}
\label{Intro} 

The thermodynamic dislocation theory (TDT), proposed initially by Langer, Bouchbinder, and Lookmann \cite{LBL-10} and developed further in \cite{JSL-15,JSL-16,JSL-17,JSL-17a,Le17,Le18}, deals with the uniform plastic deformations of crystals driven by a constant strain rate. During these uniform plastic deformations the crystal may have only redundant dislocations whose resultant Burgers vector vanishes. As shown in \cite{Le18a,LP18}, the extension of TDT to non-uniform plastic deformations should account for excess dislocations due to the incompatibility of the plastic distortion \cite{Nye53}. In all studies mentioned above, only the case of proportional loading in one direction with the constant strain rate is considered. The purpose of this paper is to explore use of TDT for non-uniform plastic deformations \cite{Le18a,LP18} in modelling crystals subjected to loading, unloading, and further loading in the opposite direction in the anti-plane shear mode. Our challenge is to simulate the stress strain curves exhibiting the Bauschinger effect and explain these based on the physical mechanism of movement and annihilation of excess dislocations during the load reversal.   

The thermodynamic dislocation theory is based on two unconventional ideas. The first of these is that, under nonequilibrium conditions, the atomically slow configurational degrees of freedom of dislocated crystals are characterized by an effective disorder temperature that differs from the ordinary kinetic-vibrational temperature. Both of these temperatures are thermodynamically well defined variables whose equations of motion determine the irreversible behaviors of these systems. The second principal idea is that entanglement of dislocations is the overwhelmingly dominant cause of resistance to deformation in crystals.  These two ideas have led to successfully predictive theories of strain hardening \cite{LBL-10,JSL-15}, steady-state stresses over exceedingly wide ranges of strain rates \cite{LBL-10}, thermal softening during deformation \cite{Le17}, yielding transitions between elastic and plastic responses  \cite{JSL-16,JSL-17a}, and shear banding instabilities \cite{JSL-17,Le18}.  

We start in Sec.~\ref{EOM} with a brief annotated summary of the equations of motion to be used here.  Our focus is on the physical significance of the various parameters that occur in them.  We discuss which of these parameters are expected to be material-specific constants, independent of temperature and strain rate, and thus to be key ingredients of the theory. In Sec.~\ref{NI} we discretize the obtained system of governing equations and develop the numerical method for its solution.  The results of numerical simulations as well as the physical explanation of the Bauschinger effect are presented in Sec.~\ref{RE}.  We conclude in Sec.\ref{CONCLUSIONS} with some remarks about the significance of these calculations.
 
\section{Equations of Motion}
\label{EOM}

Suppose, for simplicity, that the single crystal beam has a rectangular cross section, of width $c$ and height $h$, that lies in the $(x,y)$-plane. This crystal beam is placed in a ``hard device'' such that at its side boundary the displacement in the $z$-direction is prescribed: $w=\gamma (t)y$, with $\gamma (t)$ being the shear strain regarded as a control parameter. Such a hard device models the grain boundary that does not allow dislocations to reach it. If $h\gg c$, we may neglect the end-effect due to the almost translational invariance in the $y$-direction and assume that all quantities depend only on the spatial coordinate $x$.

Now, let this system be driven at a constant shear rate $\dot\gamma \equiv q_0/t_0$, where $t_0$ is a characteristic microscopic time scale. Because the system is undergoing steady-state anti-plane shear, we can replace the time $t$ by the accumulated overall shear strain $\gamma$ so that $t_0\,\partial/\partial t \to q_0\,\partial/\partial \gamma$. The equation of motion for the flow stress becomes
\begin{equation}
\label{tauydot}
\frac{\partial \tau_Y}{\partial \gamma} = \mu\,\left[1 - \frac{q(\gamma )}{q_0}\right].
\end{equation}
Note that, for the uniform plastic deformation involving only redundant dislocations $q(\gamma )/t_0$ equals the plastic shear rate $\dot{\beta}$, with $\beta $ being the uniform plastic distortion, but in general when $\beta $ is non-uniform it is not necessarily so.

The state variables that describe this system are the elastic strain $\gamma-\beta$, the areal densities of redundant dislocations $\rho^r$ and excess dislocations $\rho^g\equiv  |\beta _{,x}|/ b$ (where $b$ is the length of the Burgers vector), and the effective disorder temperature $\chi$ (cf. \cite{Kroener1992,JSL-16}). All four quantities, $\gamma-\beta$, $\rho^r$, $\rho^g$, and $\chi$, are functions of $\gamma$. 

The central, dislocation-specific ingredient of this analysis is the thermally activated depinning formula for $q$ as a function of a flow stress $\tau_Y$ and a total dislocation density $\rho=\rho^r+\rho^g$:  
\begin{align}
\label{qdef}
q(\tau_Y,\rho)&= b\sqrt{\rho} [f_P(\tau_Y,\rho)-f_P(-\tau_Y,\rho)], 
\\
f_P(\tau_Y,\rho)&=\exp\,\Bigl[-\,\frac{1}{\theta}\,e^{-\tau_Y/\tau_T(\rho)}\Bigr]. \notag
\end{align}
This is an Orowan relation of the form $q = \rho\,b\,v\,t_0$ in which the speed of the dislocations $v$ is given by the distance between them multiplied by the rate at which they are depinned from each other. That rate is approximated here by the activation terms $f_P(\tau_Y)$ and $-f_P(-\tau_Y)$, in which the energy barrier $e_P=k_BT_P$ (implicit in the scaling of $\theta=T/T_P$) is reduced by the stress dependent factor $e^{-\tau_Y/\tau_T(\rho)}$, where  $\tau_T(\rho)= \mu_T\,b \sqrt{\rho}$ is the Taylor stress. Note that, when dealing with the load reversal, antisymmetry is required in Eq.~(\ref{qdef}) both to preserve reflection symmetry, and to satisfy the second-law requirement that the energy dissipation rate, $\tau_Yq/q_0$, is non-negative.  

The pinning energy $e_P$ is large, of the order of electron volts, so that $\theta$ is very small.  As a result, $q(\tau_Y,\rho)$ is an extremely rapidly varying function of $\tau_Y$ and $\theta$.  This strongly nonlinear behavior is the key to understanding yielding transitions and shear banding as well as many other important features of crystal plasticity.  For example, the extremely slow variation of the steady-state flow stress as a function of strain rate discussed in \cite{LBL-10} is the converse of the extremely rapid variation of $q$ as a function of $\tau_Y$ in Eq.(\ref{qdef}).  

The equation of motion for the total dislocation density $\rho=\rho^r+\rho^g$ describes energy flow. It says that some fraction of the power delivered to the system by external driving is converted into the energy of dislocations, and that that energy is dissipated according to a detailed-balance analysis involving the effective temperature $\chi$.  This equation is: 
\begin{equation}
\label{rhodot}
\frac{\partial \rho}{\partial \gamma} = K_\rho \,\frac{\tau_Y\,q}{a^2\nu(\theta,\rho,q_0)^2\,\mu\,q_0}\, \Bigl[1 -\frac{\rho}{\rho_{ss}(\chi)} \Bigr],
\end{equation}
with $\rho_{ss}(\chi) =(1/a^2)e^{- e_D/\chi}$ being the steady-state value of $\rho$ at given $\chi$, $e_D$ a characteristic formation energy for dislocations, and $a$ denoting the average spacing between dislocations in the limit of infinite  $\chi$ ($a$ is a length of the order of tens of atomic spacings). The coefficient $K_\rho $ is an energy conversion factor that, according to arguments presented in  \cite{LBL-10} and \cite{JSL-17}, should be independent of both strain rate and temperature.  The other quantity that appears in the prefactor in Eq.(\ref{rhodot}) is
\begin{equation}
\label{nudef}
\nu(\theta,\rho,q_0) \equiv \ln\Bigl(\frac{1}{\theta}\Bigr) - \ln\Bigl[\ln\Bigl(\frac{b\sqrt{\rho}}{q_0}\Bigr)\Bigr].
\end{equation}

The equation of motion for the effective temperature $\chi$ is a statement of the first law of thermodynamics for the configurational subsystem: 
\begin{equation}
\label{chidot}
\frac{\partial \chi }{\partial \gamma} = K\,\frac{\tau_Y e_D \,q}{\mu\,q_0}\,\Bigl( 1 -\frac{\chi}{\chi_0} \Bigr). 
\end{equation}
Here, $\chi_0$ is the steady-state value of $\chi$ for strain rates appreciably smaller than inverse atomic relaxation times, i.e. much smaller than $t_0^{-1}$. The dimensionless factor $K$ is inversely proportional to the effective specific heat $c_{e\!f\!f}$. Unlike $K_\rho$, there is no  reason to believe that $K$ is a rate-independent constant.  In \cite{JSL-17a}, $K$ for copper was found to decrease from $17$ to $12$ when the strain rate increased by a factor of $10^6$.  Since we shall consider changes in strain rate of at most a factor of $10^2$ here, we shall assume that $K$ is a constant.  

The equation for the plastic distortion $\beta$ reads 
\begin{equation}
\label{microforces}
\tau -\tau_B -\tau_Y =0.
\end{equation} 
This equation is the balance of microforces acting on dislocations. Here, the first term $\tau=\mu(\gamma -\beta)$ is the applied shear stress, the second term the back-stress due to the interaction of excess dislocations, and the last one the flow stress. This balance of microforces can be derived from the variational equation for irreversible processes \cite{Le18a,LP18} yielding
\begin{equation}
\label{zeta}
\tau_B =-\frac{1}{b}(\frac{\partial \phi_m}{\partial \rho^g})_{,x} \text{sign} \beta_{,x}=-\frac{1}{b^2}\frac{\partial ^2\phi_m}{\partial (\rho^g)^2}\beta_{,xx},
\end{equation}
with $\phi_m$ being the free energy density of excess dislocations. Berdichevsky \cite{VB17} has found $\phi_m$ for the locally periodic arrangement of excess screw dislocations. However, as shown by us in \cite{LP18}, his expression must be extrapolated to the extremely small or large dislocation densities to guarantee the existence of solution within TDT. Using the extrapolated energy proposed in \cite{LP18} we find that
\begin{equation}
\label{backstress}
\tau_B=-\mu b^2\frac{k_1\xi^2+(2k_0k_1-1)\xi+k_1k_0^2-2k_0}{4\pi(k_0+\xi)^2}\beta_{,xx},
\end{equation}
where $\xi=b|\beta_{,x}|$. Equation (\ref{microforces}) is subjected to the Dirichlet boundary condition $\beta(0)=\beta(c)=0$.     

\begin{figure}[t]
	\centering
	\includegraphics[width=.45\textwidth]{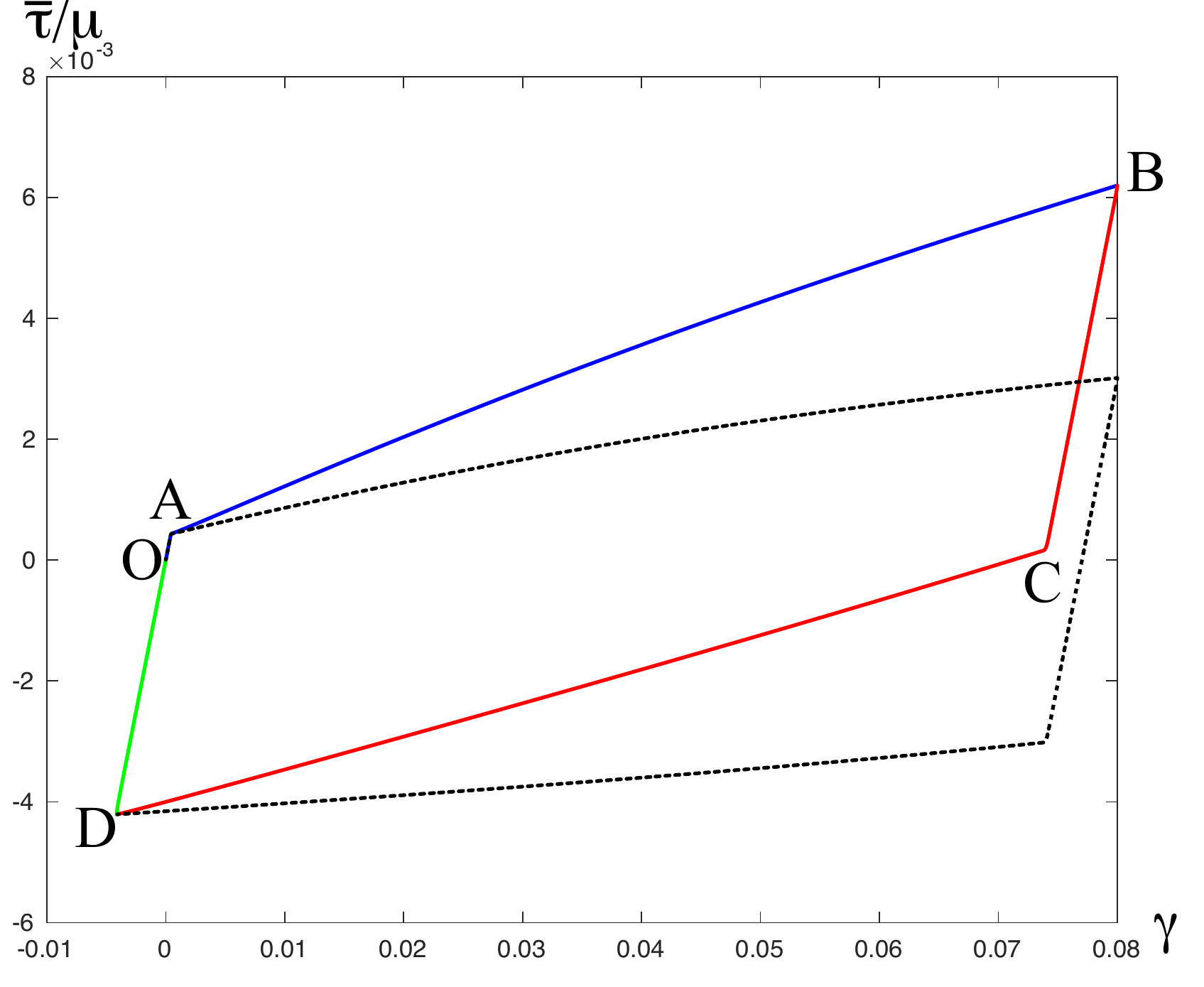}
	\caption{(Color online) Stress-strain curves at the strain rate $\tilde{q}_0=10^{-13}$, for room temperature, and for $\gamma^*=0.08$: (i) loading path OAB (blue), (ii) load reversal BCD (red), (iii) second load reversal DO (green), (iv) flow stress versus strain (dashed black curve).}
	\label{fig:1}
\end{figure}

\section{Discretization and method of solution}
\label{NI}
For the purpose of numerical integration of the system of equations (\ref{tauydot})-(\ref{backstress}) it is convenient to introduce the dimensionless variables and quantities
\begin{equation}
\label{dimless}
\tilde{x}=x/b,\, \tilde{\rho}=a^2\rho,\, \tilde{\tau}=\tau/\mu, \, \tilde{\tau}_Y=\tau_Y/\mu ,\, \tilde{\tau}_B=\tau_B/\mu . 
\end{equation}
The variable $\tilde{x}$ changes from 0 to $\tilde{c}=c/b$. Then we rewrite Eq.~(\ref{qdef}) in the form
\begin{equation}
\label{tildeq}
q(\tau_Y,\rho)=\frac{b}{a}\tilde{q}(\tilde{\tau}_Y,\tilde{\rho}),
\end{equation}
where
\begin{equation}
\label{tildeqdef}
\tilde{q}(\tilde{\tau}_Y,\tilde{\rho})=\sqrt{\tilde{\rho}}[\tilde{f}_P(\tilde{\tau}_Y,\tilde{\rho})-\tilde{f}_P(-\tilde{\tau}_Y,\tilde{\rho})].
\end{equation}
We set $\tilde{\mu}_T=(b/a)\mu_T=\mu r$ and assume that $r$ is independent of temperature and strain rate. Then
\begin{equation}
\label{tildefp}
\tilde{f}_P(\tilde{\tau}_Y,\tilde{\rho})=\exp\,\Bigl[-\,\frac{1}{\theta}\,e^{-\tilde{\tau}_Y/(r\sqrt{\tilde{\rho }})}\Bigr].
\end{equation}
We define $\tilde{q}_0=(a/b)q_0$ so that $q/q_0=\tilde{q}/\tilde{q}_0$. Eq.~(\ref{nudef}) becomes
\begin{equation}
\label{nudef1}
\tilde{\nu}(\theta,\tilde{\rho},\tilde{q}_0) \equiv \ln\Bigl(\frac{1}{\theta}\Bigr) - \ln\Bigl[\ln\Bigl(\frac{\sqrt{\tilde{\rho}}}{\tilde{q}_0}\Bigr)\Bigr].
\end{equation}
The dimensionless steady-state quantities are
\begin{equation}
\label{ss}
\tilde{\rho}_{ss}(\tilde{\chi})=e^{-1/\tilde{\chi}}, \quad \tilde{\chi}_0=\chi_0/e_D.
\end{equation}
Using $\tilde{q}$ instead of $q$ as the dimensionless measure of plastic
strain rate means that we are effectively rescaling $t_0$ by a
factor $b/a$. For purposes of this analysis, we assume that $(a/b)t_0=10^{12}$s.

\begin{figure}[t]
	\centering
	\includegraphics[width=.45\textwidth]{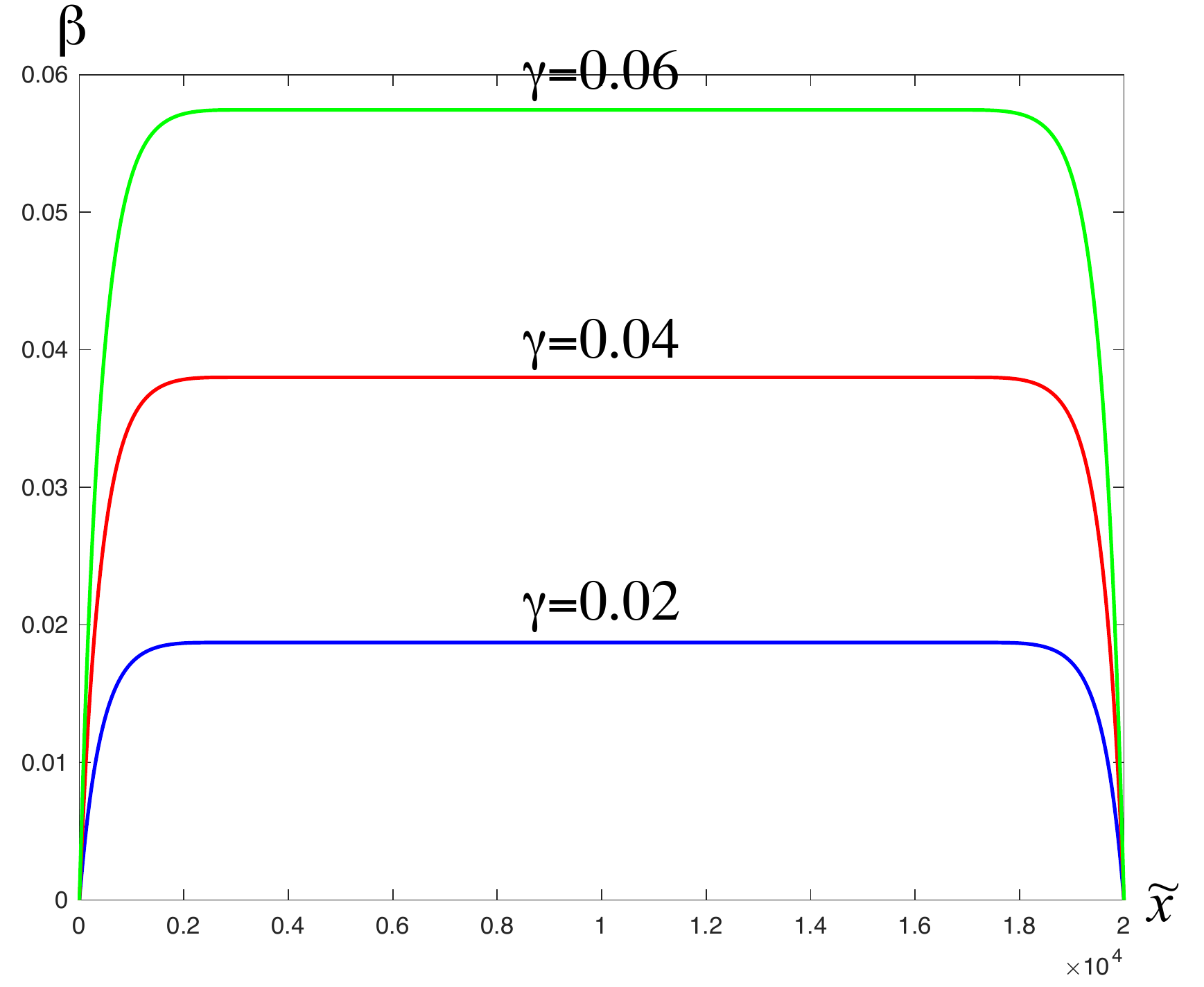}
	\caption{(Color online) Evolution of $\beta(\tilde{x})$ at the strain rate $\tilde{q}_0=10^{-13}$ and for room temperature during the loading along AB: (i) $\gamma =0.02$ (blue), (ii) $\gamma =0.04$ (red), (iii) $\gamma =0.06$ (green).}
	\label{fig:1a}
\end{figure}

In terms of the introduced dimensionless quantities the governing equations read
\begin{equation}
\frac{\partial \tilde{\tau}_Y}{\partial \gamma} = \left[1 - \frac{\tilde{q}(\tilde{\tau}_Y,\tilde{\rho})}{\tilde{q}_0}\right], \label{tau} 
\end{equation}
\begin{eqnarray}
\frac{\partial \tilde{\rho}}{\partial \gamma} = K_\rho \,\frac{\tilde{\tau}_Y\,\tilde{q}}{\tilde{\nu}(\theta,\tilde{\rho},\tilde{q}_0)^2\,\tilde{q}_0}\, \Bigl[1 -\frac{\tilde{\rho}}{\tilde{\rho}_{ss}(\tilde{\chi})} \Bigr], \\
\frac{\partial \tilde{\chi }}{\partial \gamma} = K\,\frac{\tilde{\tau}_Y\,\tilde{q}}{\tilde{q}_0}\,\Bigl( 1 -\frac{\tilde{\chi}}{\tilde{\chi}_0} \Bigr), 
\\ 
\gamma-\beta +\tilde{\tau}_B -\tilde{\tau}_Y =0. \label{force}
\end{eqnarray}
where
\begin{equation}
\label{g}
\tilde{\tau}_B =\frac{k_1\xi^2+(2k_0k_1-1)\xi+k_1k_0^2-2k_0}{4\pi(k_0+\xi)^2}\beta_{,\tilde{x}\tilde{x}},
\end{equation}
and $\xi=|\beta_{,\tilde{x}}|$. To solve this system of partial differential equations subject to initial and boundary conditions numerically, we discretize the equations in the interval $(0< \tilde x <\tilde{c})$ by dividing it into $n$ sub-intervals of equal length $\Delta \tilde{x}=\tilde{c}/n$. Then the first and second spatial derivative of $\beta$ in equation (\ref{force}) are approximated by
\begin{eqnarray}
\frac{\partial \beta}{\partial \tilde{x}}(\tilde{x}_i)=\frac{\beta_{i+1}-\beta_i}{\Delta \tilde{x}},
\\
\frac{\partial ^2\beta}{\partial \tilde{x}^2}(\tilde{x}_i)=\frac{\beta_{i+1}-2\beta_i+\beta_{i-1}}{(\Delta \tilde{x})^2},
\end{eqnarray}
where $\beta_i=\beta(\tilde{x}_i)$. In this way,  we reduce the four partial differential equations to a system of $4n$ ordinary differential-algebraic equations. We have solved these numerically using the Matlab-ode15s solver with $n =1000$ and the $\gamma$ step equal to $10^{-6}$.
 
\begin{figure}[t]
	\centering
	\includegraphics[width=.45\textwidth]{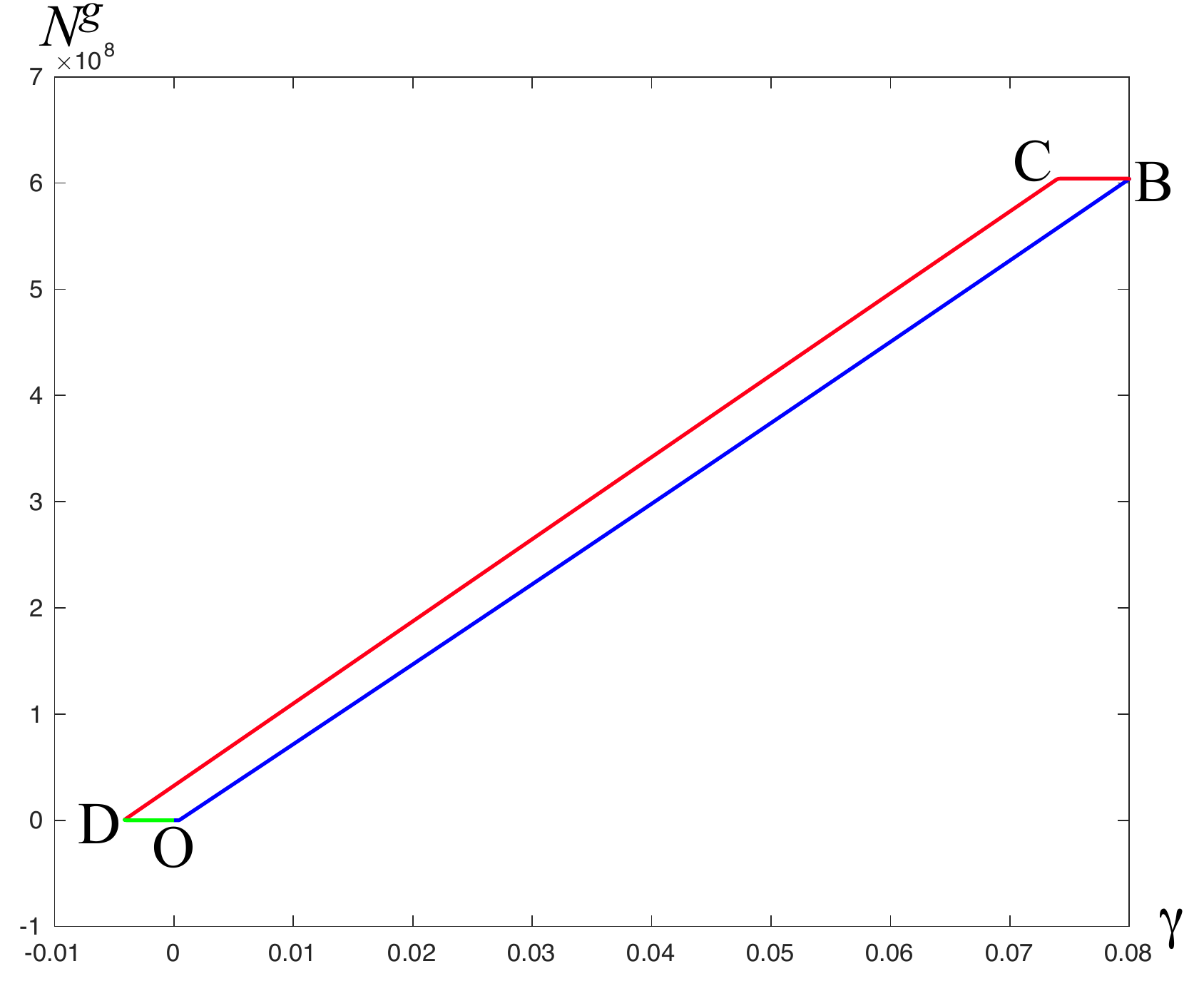}
	\caption{(Color online) $N^g$ versus $\gamma$ at the strain rate $\tilde{q}_0=10^{-13}$, for room temperature, and for $\gamma^*=0.08$: (i) loading path (blue), (ii) load reversal (red), (iii) second load reversal (green).}
	\label{fig:2}
\end{figure}

After shearing the specimen up to the shear strain $\gamma^*$ we unload the crystal and load it further in the opposite direction by reversing the direction of change of $\gamma $. The latter will now be reduced with the same rate from $\gamma^*$ to some negative value $\gamma_*$, to be specified later. We postulate that the system of governing equations (\ref{tau})-(\ref{force}) remains valid during this load reversal. Besides, as  initial conditions we propose that all quantities $\tilde{\tau}_Y$, $\tilde{\rho}$, $\tilde{\chi}$, and $\beta$ assume those values $\tilde{\tau}_Y(\gamma^*)$, $\tilde{\rho}(\gamma^*)$, $\tilde{\chi}(\gamma^*)$, and $\beta(\gamma^*)$ achieved at the end of the loading, thus satisfying the continuity requirement. Since the equations remain valid during the load reversal with the same magnitude of the strain rate, we let $\tilde{q}_0$ unchanged, reverse the expression for $\tilde{q}$ to
\begin{equation}
\label{qbar}
\bar{q}(\tilde{\tau}_Y,\tilde{\rho})=\sqrt{\tilde{\rho}}[\tilde{f}_P(-\tilde{\tau}_Y,\tilde{\rho})-\tilde{f}_P(\tilde{\tau}_Y,\tilde{\rho})],
\end{equation}
and integrate the system (\ref{tau})-(\ref{force}), with $\tilde{q}$ being replaced by $\bar{q}$, backwards in $\gamma$. If the next load reversal with the same shear rate should be made after reaching $\gamma_*$, we switch again to $\tilde{q}$ and integrate the system (\ref{tau})-(\ref{force}), with the initial conditions satisfying the continuity requirement, forwards in $\gamma$. We can also change the magnitude of the strain rate $\tilde{q}_0$ if necessary. Thus, the numerical simulation with several load reversals and different strain rates can be realized in this way.

\begin{figure}[t]
	\centering
	\includegraphics[width=.45\textwidth]{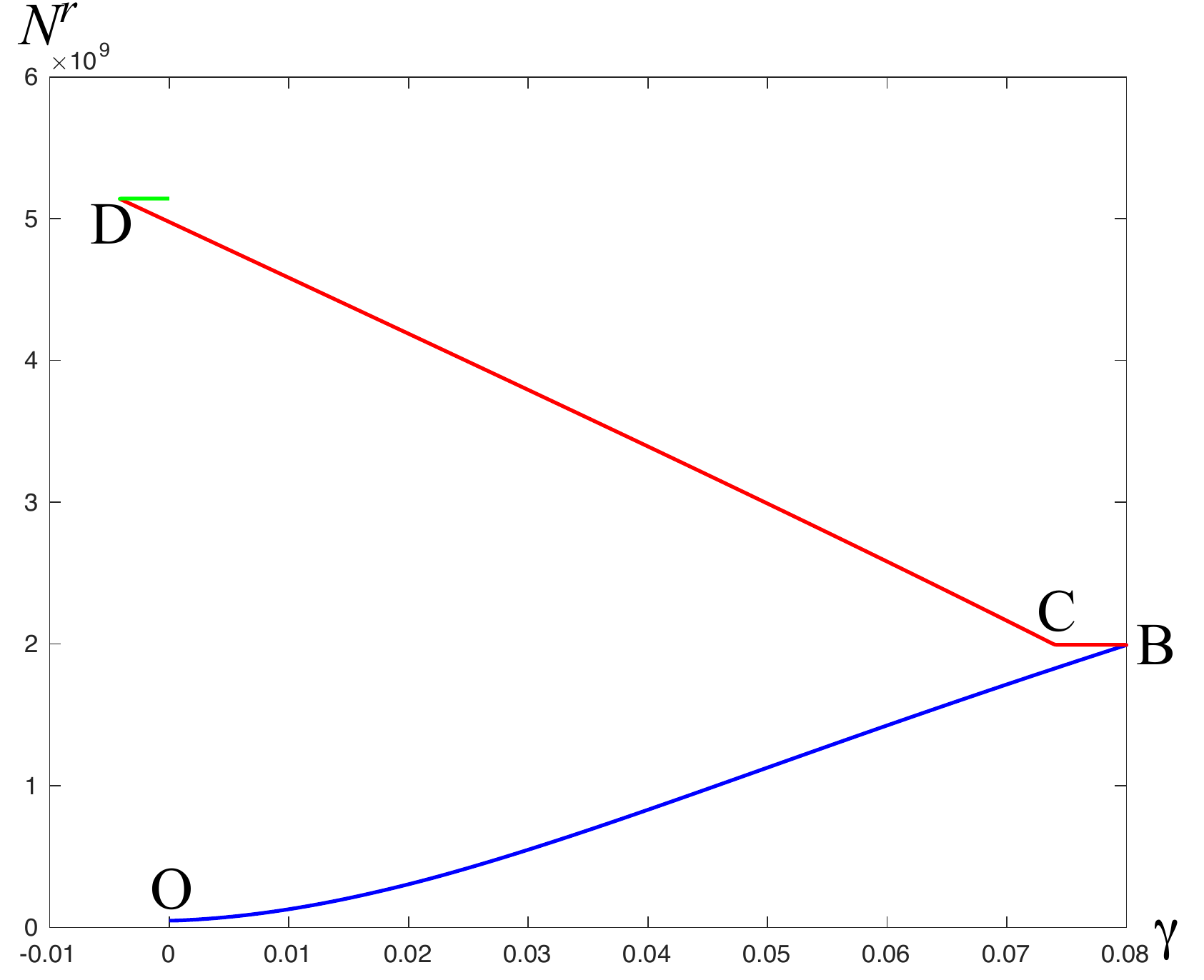}
	\caption{(Color online) $N^r$ versus $\gamma$ at the strain rate $\tilde{q}_0=10^{-13}$, for room temperature, and for $\gamma^*=0.08$: (i) loading path (blue), (ii) load reversal (red), (iii) second load reversal (green).}
	\label{fig:3}
\end{figure}

After finding the solution we can compute the average rescaled stress according to
\begin{equation}
\label{avstress}
\bar{\tau}/\mu =\frac{1}{\tilde{c}} \int_0^{\tilde{c}} \tilde{\tau} d\tilde{x}.
\end{equation}
The average flow stress is computed similarly. The total number of dislocations per unit height is
\begin{equation}
\label{totalN}
N=\int_0^c \rho \, dx=\frac{b}{a^2} \int_0^{\tilde{c}} \tilde{\rho }\, d\tilde{x}.
\end{equation}
The number of excess dislocations per unit height equals
\begin{equation}
\label{excessN}
N^g=2\int_0^{c/2} \rho ^g\, dx=\frac{2}{b}\int_0^{c/2} \beta_{,x}\, dx=\frac{2}{b}\beta_m,
\end{equation}
where $\beta_m=\beta(\tilde{c}/2)$. Then, obviously, $N^r=N-N^g$. 

\section{Numerical simulations}
\label{RE}

\begin{figure}[htp]
	\centering
	\includegraphics[width=.45\textwidth]{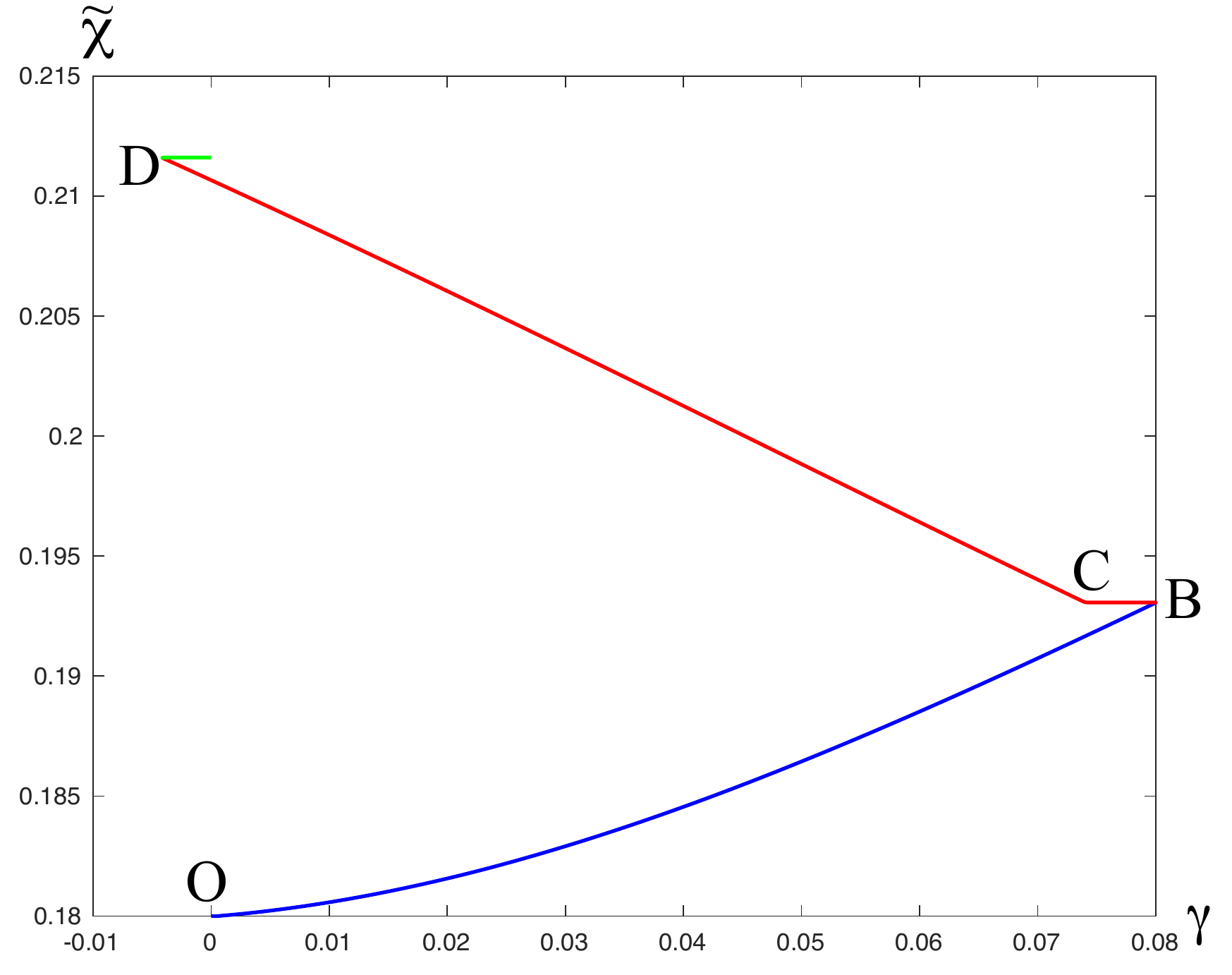}
	\caption{(Color online) $\tilde{\chi}(\tilde{c}/2)$ versus $\gamma$ at the strain rate $\tilde{q}_0=10^{-13}$, for room temperature, and for $\gamma^*=0.08$: (i) loading path (blue), (ii) load reversal (red), (iii) second load reversal (green).}
	\label{fig:4}
\end{figure}

Let the anti-plane shear test be done at room temperature $T=298$K and at the strain rate $\tilde{q}_0=10^{-13}$. The parameters for copper at this room temperature are chosen as follows \cite{LBL-10} 
\begin{equation*}
r=0.0323, \, \theta =0.0073, \, K=350, \, K_\rho =96.1,\, \tilde{\chi }_0=0.25.
\end{equation*}
We take $c=5.1$ micron, $b=2.55$\AA \, and $a=10b$. In addition, the parameters $k_0$ and $k_1$ required to compute the back stress \cite{LP18} are: $k_0=10^{-6}$, $k_1=2.1\times 10^6$. We choose also the initial conditions 
\begin{equation*}
\tilde{\tau}_Y(0)=0,\, \tilde{\rho}(0)=6.25\times 10^{-5}, \, \tilde{\chi}(0)=0.18,\, \beta(0)=0.
\end{equation*}
This initial dislocation density in real dimension equals $10^{13}$/m$^2$.

\begin{figure}[htp]
	\centering
	\includegraphics[width=.45\textwidth]{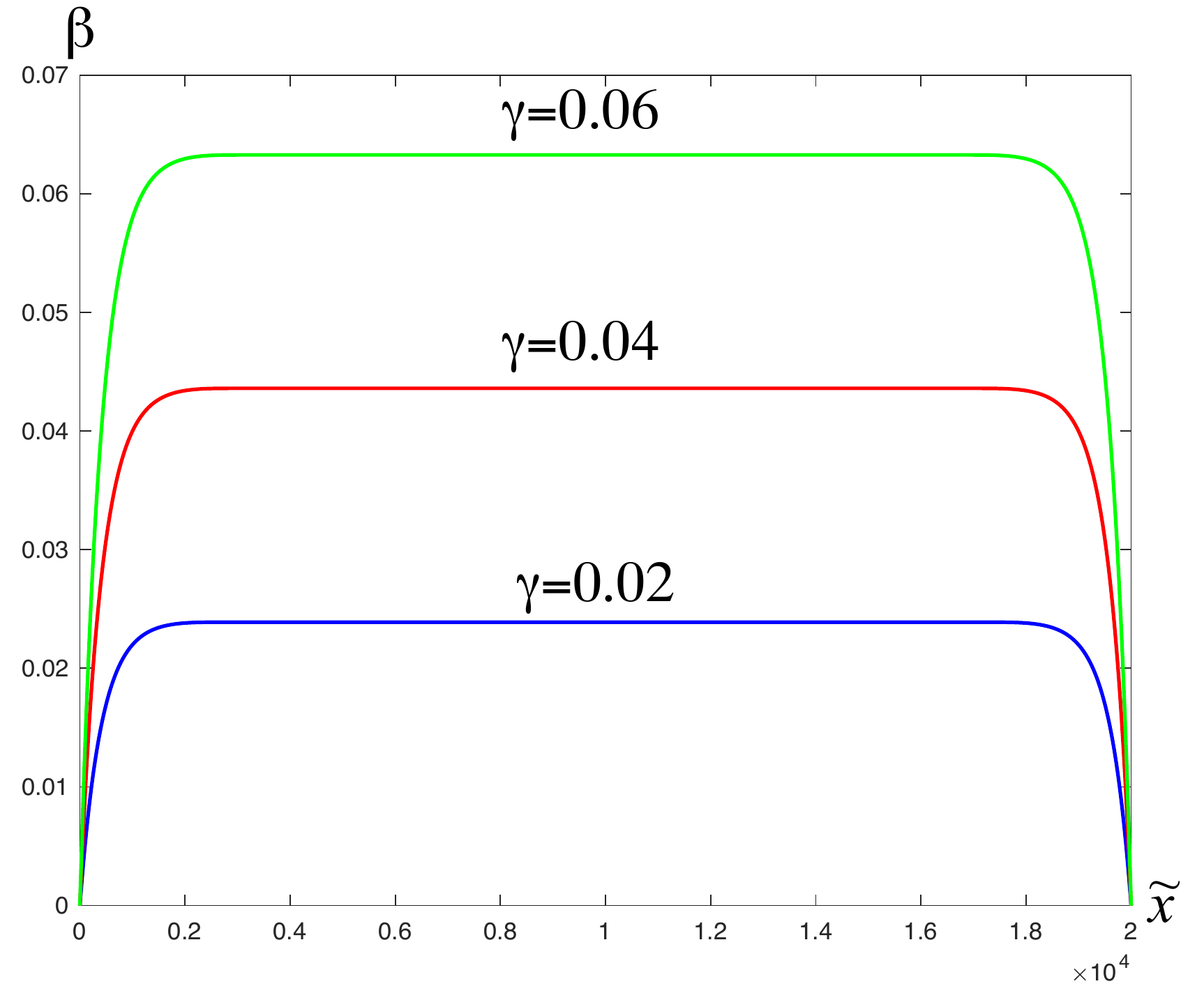}
	\caption{(Color online) Evolution of $\beta(\tilde{x})$ at the strain rate $\tilde{q}_0=10^{-13}$ and for room temperature during the load reversal along CD: (i) $\gamma =0.02$ (blue), (ii) $\gamma =0.04$ (red), (iii) $\gamma =0.06$ (green).}
	\label{fig:1b}
\end{figure}
 
\begin{figure}[htp]
	\centering
	\includegraphics[width=.45\textwidth]{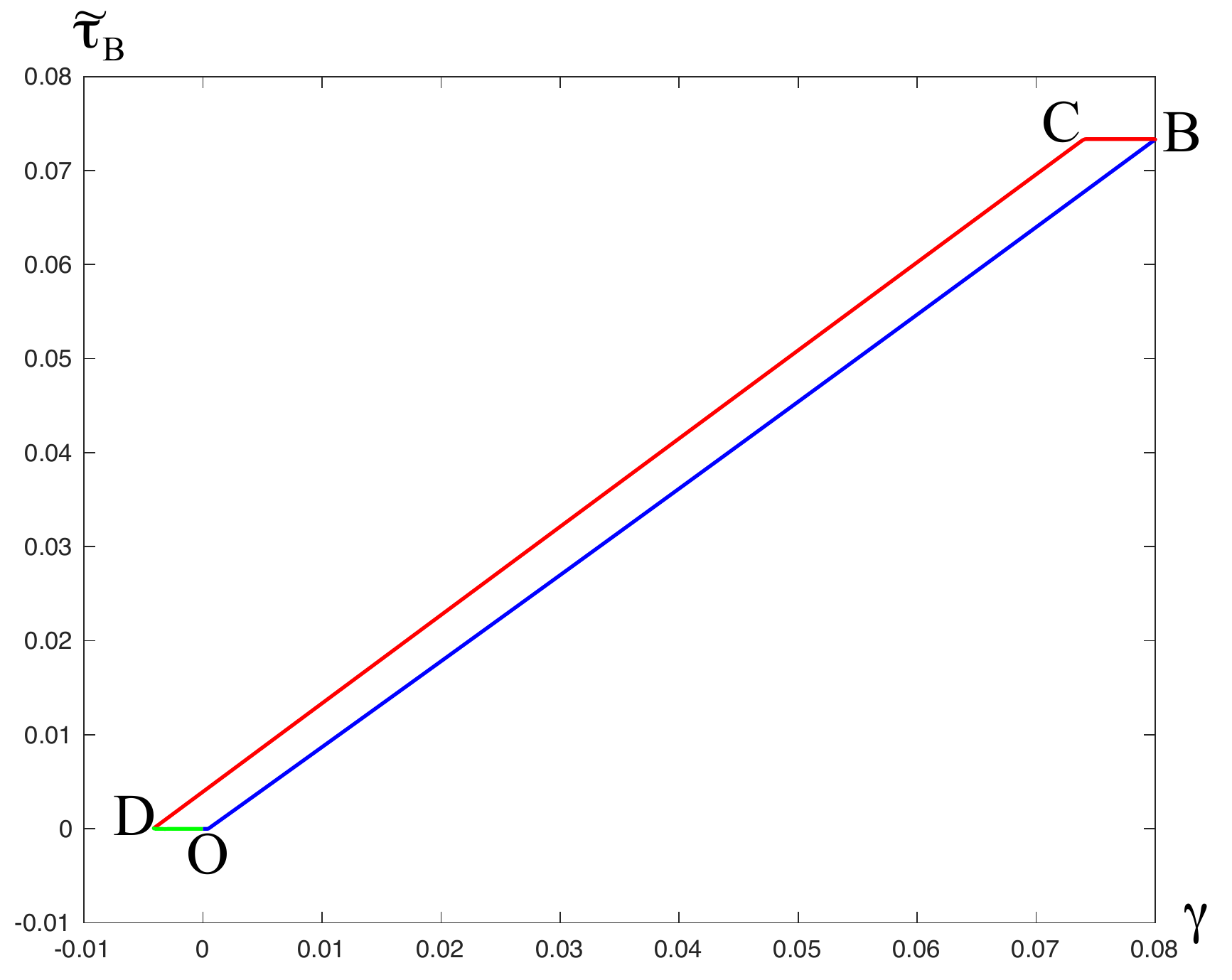}
	\caption{(Color online) The normalized back stress $\tilde{\tau}_B$ near the boundary versus $\gamma$ at the strain rate $\tilde{q}_0=10^{-13}$, for room temperature, and for $\gamma^*=0.08$: (i) loading path (blue), (ii) load reversal (red), (iii) second load reversal (green).}
	\label{fig:5}
\end{figure}

\begin{figure}[htp]
	\centering
	\includegraphics[width=.45\textwidth]{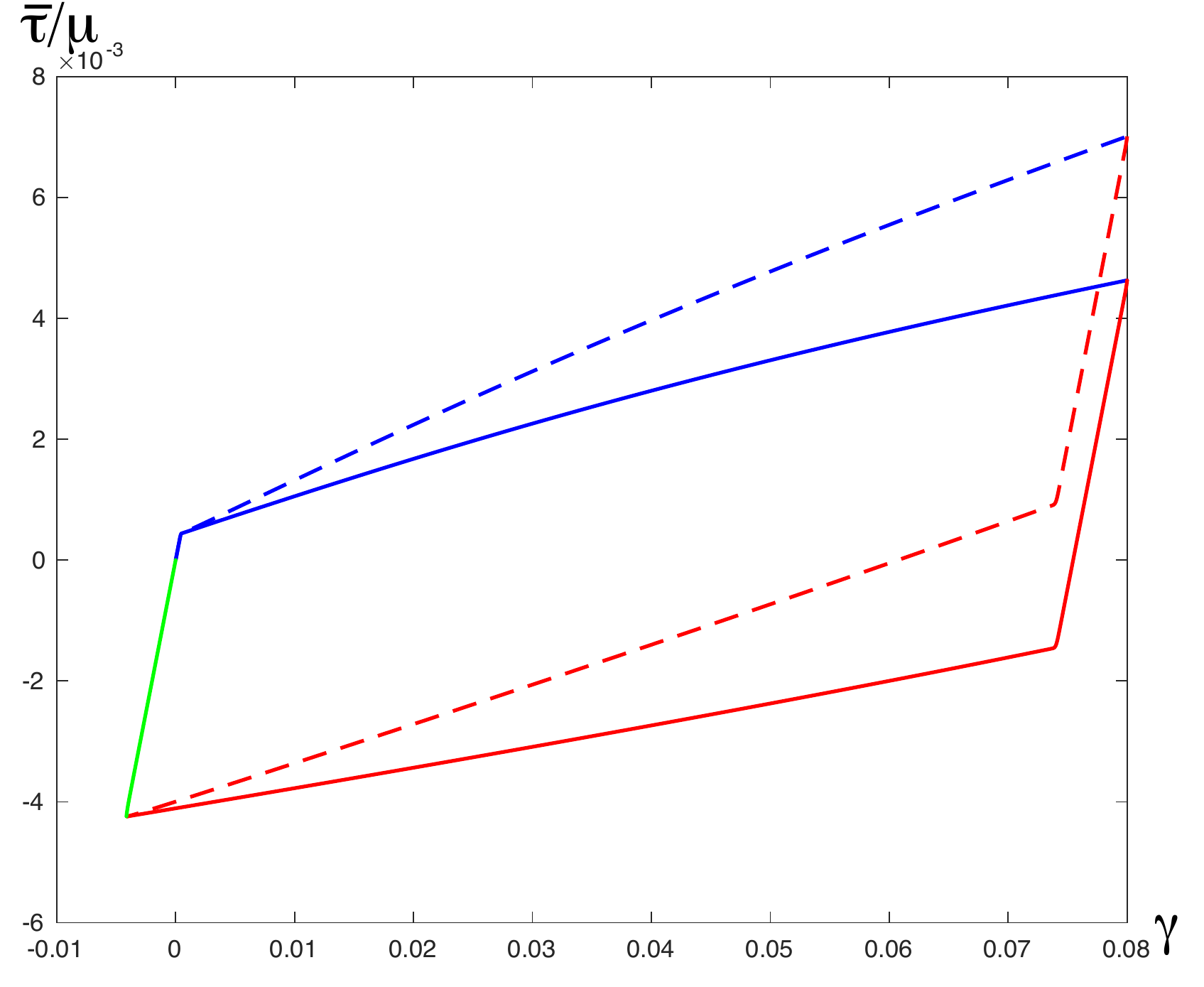}
	\caption{(Color online) The stress-strain curves for specimens with different sizes at the strain rate $\tilde{q}_0=10^{-13}$, for room temperature, and for $\gamma^*=0.08$: (i) $c=5.1$ micron (dashed), (ii) $c=51$ micro (bold).}
	\label{fig:6}
\end{figure}

\begin{figure}[htp]
	\centering
	\includegraphics[width=.45\textwidth]{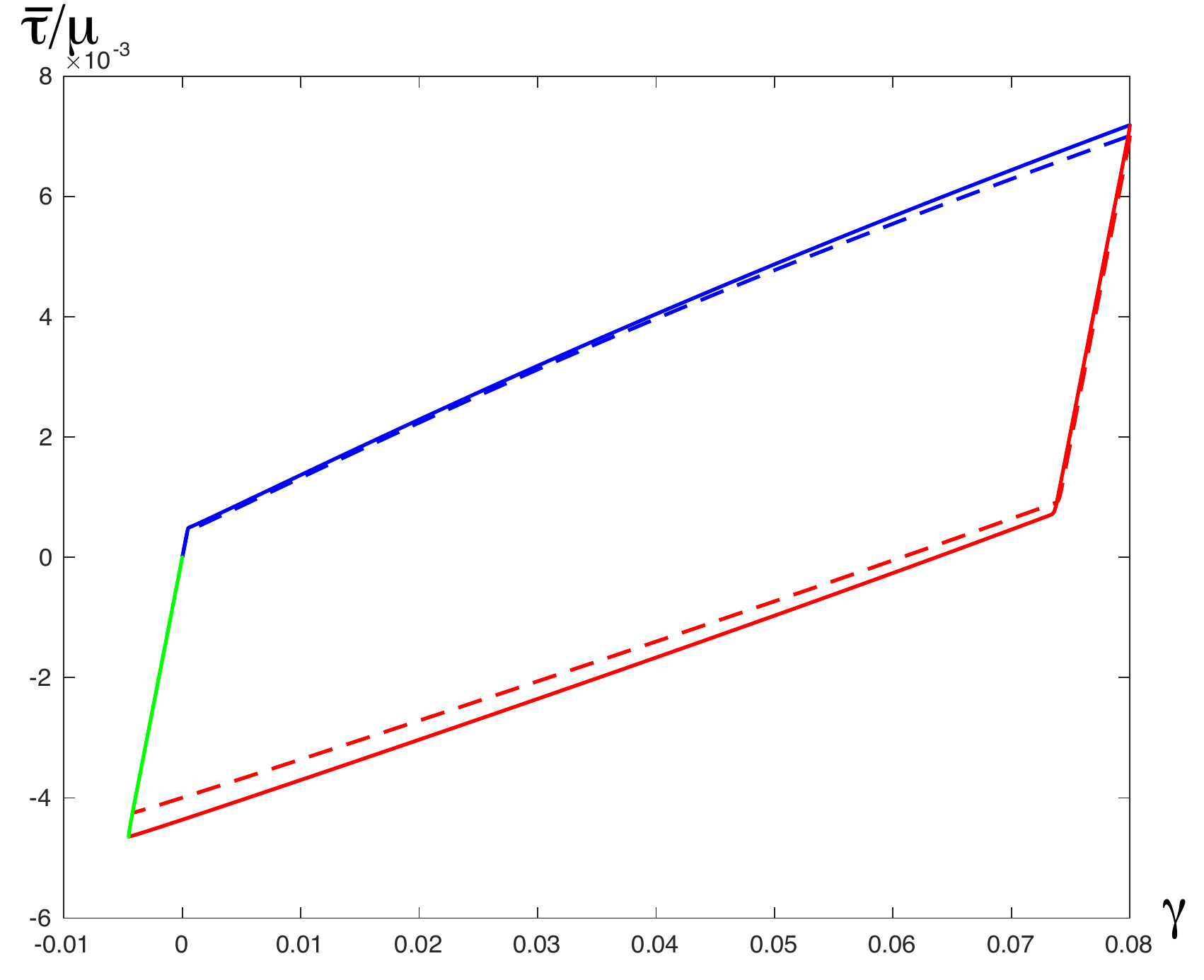}
	\caption{(Color online) The stress-strain curves for the specimen loaded at different strain rates, for room temperature, and for $\gamma^*=0.08$: (i) $\tilde{q}_0=10^{-13}$ (dashed), (ii) $\tilde{q}_0=10^{-11}$ (bold).}
	\label{fig:7}
\end{figure}

The results of numerical simulations are presented in Figs.~\ref{fig:1}-\ref{fig:7}. In Fig.~\ref{fig:1} the average normalized shear stress versus shear strain curve (called for short stress-strain curve) with the strains at the beginning of load reversals $\gamma^*=0.08$ and $\gamma_*=-0.00414$ is shown. We plot there also the average rescaled flow stress $\bar{\tau}_Y/\mu$ versus $\gamma$ (dashed black curve) for comparison. The loading path OAB (blue) consists of the elastic line OA and the hardening curve AB. The yielding transition occurs at A. Fig.~\ref{fig:1a} shows the plastic distortion $\beta $ at three different $\gamma$ along the hardening curve AB that agrees well with the approximate analytical solution found in \cite{LP18}. The excess dislocations pile up against the left and right boundaries, leaving the center of the specimen free of excess dislocations. As $\gamma $ increases the number of excess and redundant dislocations as well as the effective temperature also increase as shown in Fig.~\ref{fig:2}-\ref{fig:4}.

During the load reversal BCD (red) we observe first the elastic unloading BC where the redundant and excess dislocations as well as the effective temperature (in the middle of the specimen) are frozen as seen in Figs.~\ref{fig:2}-\ref{fig:4}. The yielding transition occurs at C, where the magnitude of the stress is much lower than that at the end of the loading path exhibiting the Bauschinger effect. To explain this effect we plot in Fig.~\ref{fig:5} the evolution of the normalized back stress $\tilde{\tau}_B$ (near the boundary) as $\gamma $ changes. This back stress increases during the loading due to the increasing number of excess dislocations, and then remain unchanged during the unloading when the dislocations are frozen along the line BC. The presence of this positive back stress reduces the magnitude of shear stress required for pulling the excess dislocations back to the center of the specimen. There, the excess dislocations of opposite signs meet and annihilate each other, so the number of excess dislocations reduces gradually to zero along the curve CD as shown in Fig.~\ref{fig:2}. Fig.~\ref{fig:1b} shows the evolution of the plastic distortion at three different $\gamma$ that confirms this tendency. It is interesting that the number of redundant dislocations do not decrease at all, except that they are also frozen along the elastic line BC as shown in Fig.~\ref{fig:3}. 

If we reverse the loading direction again by increasing $\gamma $ from $\gamma_*$ to zero, $(\gamma,\tilde{\tau})$ moves along the elastic line DO where the dislocations and the effective temperature are frozen (see Fig.~1). Note that the effective temperature always increases during the loading along AB and loading in the opposite direction along CD as seen in Fig.~\ref{fig:4}, in agreement with the second law of thermodynamics. 

It is remarkable that the loading path in the opposite direction CD differs essentially from the loading path AB due to the increase of the total number of dislocations along CD. Thus, along CD the material is closer to the steady state than along AB, and consequently, the slope of CD must be less than that of AB. This asymmetry between loadings in opposite directions becomes more pronounced as $\gamma^*$ increases.

It is interesting to examine the influence of the size of the sample on the Bauschinger effect. Fig.~\ref{fig:6} shows two stress-strain curves for two samples with different thicknesses $ c =5.1$ micron (dashed line) and $c = 51$ micron (bold line) and with all other parameters being left unchanged. We see that the size strongly influences the slope of the hardening curve because the dislocation pile-up for the smaller sample leads to stronger kinematic hardening than for the larger sample (smaller is stronger). This affects the stress level at which the yielding transition occurs during load reversal. For the smaller sample, this stress is even positive.

Another important question is how much the strain rate affects the Bauschinger effect. Fig.~\ref{fig:7} shows two stress-strain curves for two samples loaded at two different strain rates $ \tilde {q}_0 = 10^ {-13} $ (dashed line) and $ \tilde{q}_0 = 10^{-11} $ (bold line). The thickness of both samples is $ c = 5.1$ micron, while all other parameters remain unchanged. We see that the strain rate mainly affects the isotropic hardening, but much less the Bauschinger effect. The reason is that the kinematic hardening due to the excess dislocations is much less sensitive to the change in strain rate. Also note that it is practically difficult to stop the load and instantly realize the load reversal at the same rate of strain, especially for the fast loading. Therefore, the real stress-strain curve usually deviates from the theoretical one in a small transient period. Perhaps this is one of the reasons for the differences between unloading and immediate reloading.

\section{Concluding Remarks}
\label{CONCLUSIONS}
  
The physical explanation of the Bauschinger effect on the basis of the back stress and the excess dislocations seems to us to be quite reasonable. In constructing the STZ-theory for glasses \cite{JSL-08} Langer has argued that ``The Bauschinger effect is one example where the system remembers the direction in which it has been deformed, and responds differently -- more compliantly or less so -- to further loading in different directions. The natural way to include such effects in the theory is to let the STZ’s possess internal degrees of freedom that carry information from one event to the next.'' This argument holds true for dislocation mediated plasticity as well. Here, in our opinion, the incompatible plastic distortion is the natural variable that keeps the memory of excess dislocations. It cannot enter the free energy, but the curl of this quantity should enter the free energy causing the back stress. In this way the theory differs substantially from the phenomenological plasticity that introduces the back stress along with an assumed constitutive equation to fit the stress strain curves exhibiting the kinematic hardening. In contrary, our theory allows us to find the back stress from the first principle calculation of the free energy of dislocated crystals and thus to predict the stress-strain curves and the Bauschinger effect.

As the comparison with the experiments is concerned, the experimental data in anti-plane shear-controlled deformations are not known to us, in contrast to the tension-compression tests provided in \cite{Abel72,Bate86} or plain strain shear tests in \cite{Lew03,Wincze05}, so the justification of the theory by the experiments is not possible at present time. However, the propose theory may serve as a wonderful guide for the future experimental investigation on the Bauschinger effect in several directions: (i) the asymmetry between loadings in opposite directions at different level of $\gamma ^*$, (ii) the influence of the size effect, (iii) the sensitivity of the back stress on the strain rate and temperature, et cetera. Last, but not least, let us mention that the thermodynamic dislocation theory for uniform plastic deformations \cite{LBL-10} yields in the proportional compression tests for copper an excellent agreement with the experiments conducted in \cite{Follansbee1998} over a wide range of temperatures and strain rates. This gives us the hope that the same will happens for the theory proposed in this paper. 

\begin{acknowledgments}

T.M. Tran acknowledges support from the Vietnamese Government Scholarship Program 911. K.C. Le is grateful to J.S. Langer for helpful discussions.

\end{acknowledgments}

\begin{thebibliography}{99}
 

\bibitem{LBL-10} J.S. Langer, E. Bouchbinder and T. Lookman, Acta Mat. {\bf 58}, 3718 (2010).

\bibitem{JSL-15} J.S. Langer, Phys. Rev. E {\bf 92}, 032125 (2015).

\bibitem{JSL-16} J.S. Langer, Phys. Rev. E {\bf 94}, 063004 (2016).

\bibitem{JSL-17} J.S. Langer, Phys. Rev. E. {\bf 95}, 013004 (2017).

\bibitem{JSL-17a} J.S. Langer, Phys. Rev. E. {\bf 95}, 033004 (2017).

\bibitem{Le17} K.C. Le, T.M. Tran and J.S. Langer, Phys. Rev. E. {\bf 96}, 013004 (2017).

\bibitem{Le18} K.C. Le, T.M. Tran and J.S. Langer, Scripta Mat. {\bf 149}, 62 (2018).

\bibitem{Le18a} K.C. Le, J. Mech. Phys. Solids {\bf 111}, 157 (2018).

\bibitem{LP18} K.C. Le and Y. Piao, arXiv:1801.05304 (2018).

\bibitem{Nye53} J.F. Nye, Acta Metall. {\bf 1}, 153 (1953).

\bibitem{Kroener1992}
E. Kr{\"o}ner, GAMM-Mitteilungen {\bf 15}, 104 (1992). 

\bibitem{VB17} V.L. Berdichevsky, Int. J.  Eng. Sci. {\bf 116}, 74 (2017). 

\bibitem{JSL-08} J.S. Langer, Physical Review E {\bf 77}, 021502 (2008).

\bibitem{Abel72} A. Abel and H. Muir, Philos. Mag. {\bf 26}, 489 (1972). 

\bibitem{Bate86} P.S. Bate and D.W. Wilson, Acta Metall. {\bf 34} 1097 (1986).

\bibitem{Lew03} M. Lewandowska, Mat. Chem. Phys. {\bf 81} 555 (2003).

\bibitem{Wincze05} G. Vincze, E.F. Rauch, J.J Gracio, F. Barlat and A.B. Lopes, Acta Mater. {\bf 53} 1005 (2005). 

\bibitem{Follansbee1998}
P.S. Follansbee and U.F. Kocks, Acta Metall. {\bf 36}, 81 (1998).

\end{thebibliography}
\end{document}